\documentclass{emulateapj}
\usepackage{rotating}
\shorttitle{Image Family Identification with Gini Coefficient}
\shortauthors{Florian et al.}

\begin{document}

\title{The Gini Coefficient as a Tool for Image Family Idenitification in Strong Lensing Systems with Multiple Images}

\author{Michael K. Florian\altaffilmark{1,2}}

\author{Michael D. Gladders\altaffilmark{1,2}}

\author{Nan Li\altaffilmark{1,2,3}}

\author{Keren Sharon\altaffilmark{4}}

\altaffiltext{1}{Department of Astronomy and Astrophysics, University of Chicago, Chicago, IL 60637}
\altaffiltext{2}{Kavli Institute for Cosmological Physics, The University of Chicago, Chicago, IL 60637}
\altaffiltext{3}{Argonne National Laboratory, 9700 South Cass Avenue B109, Lemont, IL 60439}
\altaffiltext{4}{Department of Astronomy, University of Michigan, 1085 S. University Ave., Ann Arbor, MI 48109, USA}

\begin{abstract}
The sample of cosmological strong lensing systems has been steadily growing in recent years and with the advent of the next generation of space-based survey telescopes, the sample will reach into the thousands.  The accuracy of strong lens models relies on robust identification of multiple image families of lensed galaxies.  For the most massive lenses, often more than one background galaxy is magnified and multiply-imaged, and even in the cases of only a single lensed source, identification of counter images is not always robust.  Recently, we have shown that the Gini coefficient in space-telescope-quality imaging is a measurement of galaxy morphology that is relatively well-preserved by strong gravitational lensing.  Here, we investigate its usefulness as a diagnostic for the purposes of image family identification and show that it can remove some of the degeneracies encountered when using color as the sole diagnostic, and can do so without the need for additional observations since whenever a color is available, two Gini coefficients are as well.
\end{abstract}

\keywords{}

\section{Introduction}
The next generation of surveys, using telescopes like the Wide-Field Infrared Survey Telescope (WFIRST) and Euclid, will reveal thousands of strong-lensing galaxy clusters.  The ability to create models of these lensing masses will be crucial to fully leveraging these datasets for the study of dark matter, structure formation, cluster physics, early galaxy and star formation, cosmology, and more (e.g., \citealp{Barn03,Mahdi14,Jullo10,CLASH2012,CLASH2014}).  However, making such models is resource intensive, requiring significant computing time while also, in particular, requiring a significant investment of researcher time.  Furthermore, follow-up imaging in other bands or for longer exposure times may be required to clarify image family identifications, further consuming telescope resources.

Indeed, the identification of image families is an uncertainty of particular importance for lens modelers.  Changing which image family to which a single image belongs can result in significant changes to the resulting lens models (e.g., \citealp{Jau14,Sharon0327}).  Currently, color is one of the primary pieces of information used to distinguish members of a multiply-imaged lensing system from galaxies in the foreground.  Color has the advantage of being a property of images that can be measured algorithmically and used as a constraint on the image identifications made by various modeling teams.  With Hubble Space Telescope (HST) quality imaging, one also uses internal image morphology as a constraint, but this is typically assessed by eye and requires a substantial investment of time by an experienced lens modeler.  A number of lens models for Frontier Fields (FF) clusters  have been recently published \citep{Jau14,Jau15,Johnson14,Rich14,Grillo15,Ish15,Treu15,Kaw15} where image family identification has been done using such methods.  Clearly, though, it is impractical to obtain deep imaging in so many different filters (the FF clusters, for example, are observed in seven different filters) for each of the thousands of strong lens systems that will be discovered in the coming years.  Therefore, maximizing the strong lensing constraints available from the survey data that will exist is of particular utility.

Recently, we have found the Gini coefficient to be well-preserved by strong gravitational lensing in HST-quality imaging \citep{Flo15}.  Since it is a measurement that can be made in a single filter and in the image plane, it should be possible to gain additional constraints from the Gini coefficient (at least one per filter) to help with image family identification without using any additional observational resources.  In this letter, we use the results of the simulations presented in \citet{Flo15} to show that the Gini coefficient can indeed be used in this way.

\section{The Gini Coefficient}
The Gini coefficient was introduced to astronomy by \citet{Ab2003} and has since been used in morphological studies of unlensed galaxies (e.g. \citealt{Lotz06}).  It is a measurement of the inequality of the distribution of light in a galaxy.  Conceptually, it is calculated by ordering the pixels in a given aperture by flux, and producing the cumulative distribution function, and finding the area between that curve and the curve representing the cumulative distribution function that describes a galaxy with a perfectly flat profile.  In practice, it is calculated in the following way:
\begin{equation}
G = \frac{1}{\overline{X}n(n - 1)} \sum_{i=1}^{n} (2i - n - 1)X_{i}
\end{equation}
where $X_{i}$ is the flux of the $i^{th}$ flux-ordered pixel, $\overline{X}$ is the mean flux, and $n$ is the total number of pixels within the aperture.  A Gini coefficient of 0 indicates a perfectly uniform profile and a Gini coefficient of 1 indicates an aperture where all of the light is located in a single pixel.  For more details, see \citet{Ab2003}.

\section{The Simulated Images}
The strong lensing simulations used here are described in \citet{Flo15}.  In brief, we selected low redshift galaxies (11 elliptical, 20 spiral, and 2 irregular) from the CANDELS UDS field to be used as source galaxies for a gravitational lensing ray-tracing code.  Low redshift galaxies were chosen because direct HST observations of higher redshift galaxies do not contain as much small scale complexity due to the finite resolution of the telescope.  These source galaxies were placed at redshift $z=1$ and lensed by an analytical spherical NFW-profile with $M_{200} = 10^{15}M_{\odot}h^{-1}$ and concentration parameter $c=5$ placed at $z=0.2$.  For each of the 33 sources, images were produced for 50 unique positions in the source plane near caustics, in each of 3 filters (HST ACS/WFC F814W and F606W, and WFC3/IR F160W).  A gravitational lensing ray-tracing code \citep{Li15} was run for each of these configurations, and images of the resulting image plane configurations were produced.  These were convolved with appropriate HST-like PSFs and resampled to HST-like pixel scales (0.03 arcseconds per pixel).  Finally, Gaussian noise was added to degrade the average S/N per pixel of each arc to 0.1.  Gini coefficients and colors were measured only for the arcs that were tangential arcs or counterimages (we did not consider typically demagnified central images or radial arcs because they are often either not seen, or significantly contaminated by cluster galaxy or intracluster light).

For purposes of measuring the Gini coefficient and associated uncertainties, apertures were defined as in \citet{Flo15}, one for each filter.  Gini coefficients were calculated according to Eq. 1 and uncertainties were bootstrapped as in \citet{Ab2003}.  However, for colors, apertures were obtained from the stack of the images in the F160W and F814W filters.  The resulting aperture was then applied to the F160W image and the F814W image separately. The total flux within this aperture was measured in each filter and the results were converted to instrumental magnitudes.  Colors were defined as the difference between the magnitudes in each filter (F814W - F160W and F606W - F814W in this paper).  Uncertainties in the colors were determined by making many separate noisy realizations of each simulated arc and finding the standard deviations of the resulting distributions.  For each arc, an aperture was created.  100 different noise fields were then applied to the original image (with the average SNR remaining 0.1 per pixel), in each of the two filters.  Finally, the aperture was applied to each of these 100 realizations and the flux and magnitude were calculated.

For further details of the simulation, the aperture definitions, or the method of measuring the Gini coefficent, see \citet{Flo15}.

\section{Separation of Source Galaxies in Gini-Color Space}
\begin{sidewaysfigure*}
\centering
\includegraphics[width=1.0\textwidth]{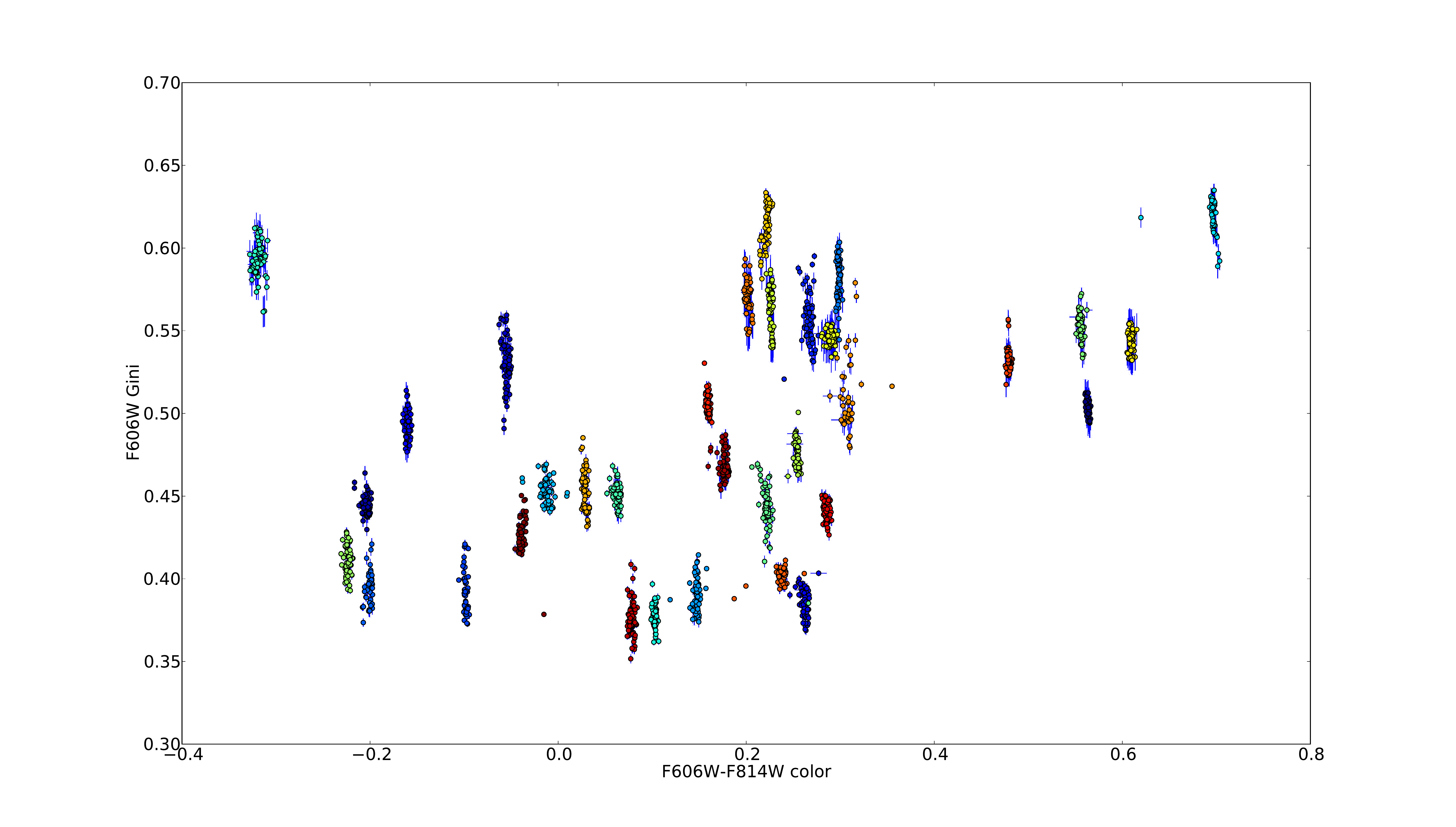}
\caption{The Gini coefficient in the F606W filter plotted against the 606W-F814W color.  The points are color-coded by source the source galaxy that was used to make each simulated arc (ie. all points of the same color are different arcs created by lensing the same source galaxy, but at different positions relative to the caustics).  Arcs from the same source galaxy tend to clump together in this space.}
    \label{Gini814vColor}

\end{sidewaysfigure*}

To determine whether the Gini coefficient can be used, along with color, to help identify the image family to which a given image belongs, we plotted, for each source, the Gini coefficient in the F606W filter against the F606W-F814W color for every lensed tangential arc and counterimage from every model realization of that galaxy.  The result is shown in Fig.~\ref{Gini814vColor}.  In this figure, the color of each point corresponds to the source galaxy (i.e., all points of a given color are different lensed images of the same source galaxy).

\begin{figure*}
\centering
\includegraphics[width=1.0\textwidth]{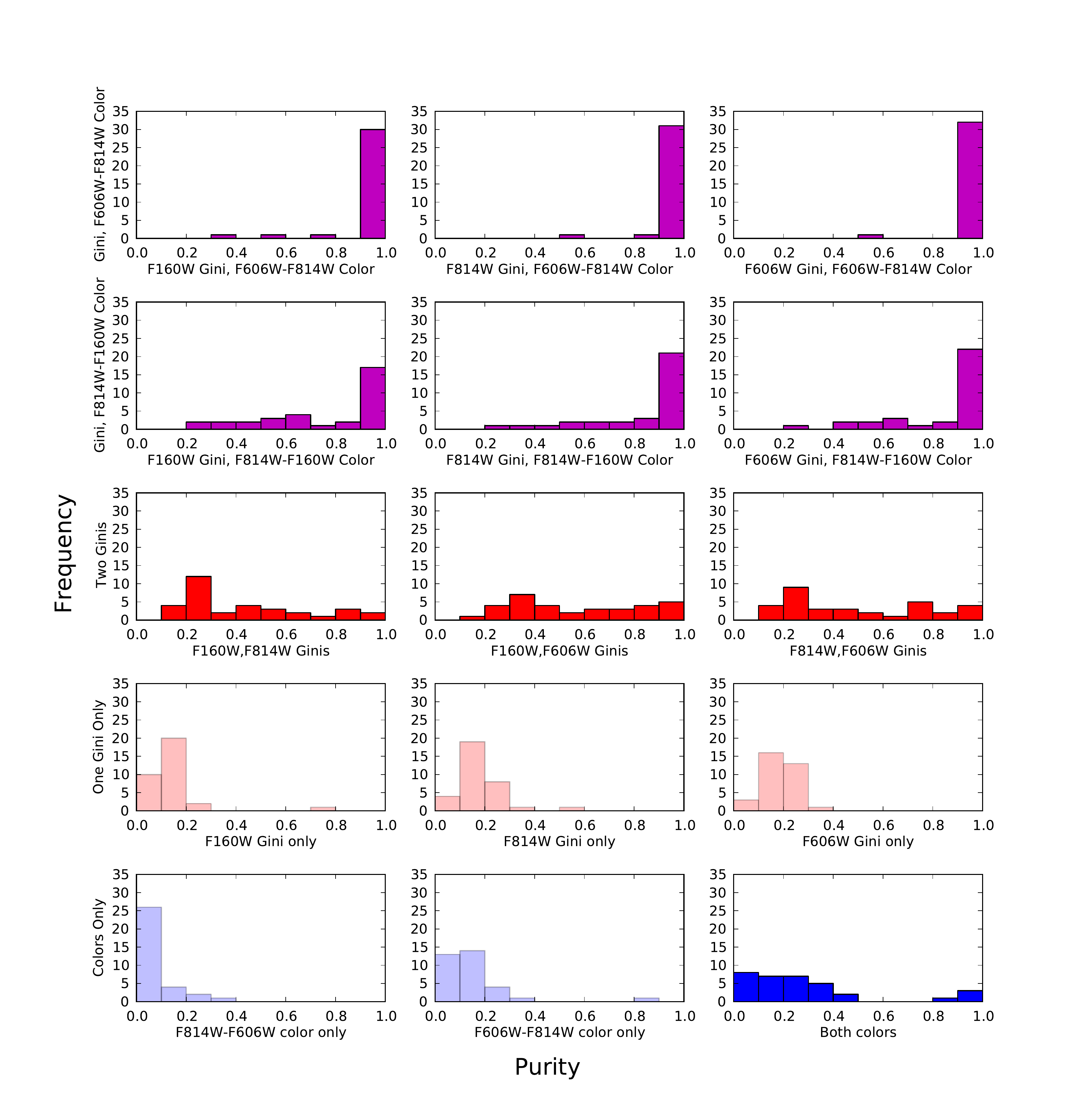}
\caption{Histograms of purities of ``clumps" like those in Fig. 1 determined from different combinations of data.  Red histograms included only Gini coefficients.  Blue histograms included only colors.  Purple histograms use a Gini coefficient and color pair.  Light red and light blue histograms use only a single color or a single Gini coefficient, while the darker ones use pairs of Ginis or pairs of colors.  Purities are the highest in general for clumps defined using one Gini coefficient and one color.}
    \label{purities}

\end{figure*}

From the figure, it is clear that different lensed images of the same source galaxy clump together in Gini-color space, suggesting that the combination of these measurements can indeed be used to distinguish between members of different image families.  It is, of course, possible that two galaxies can have the same color and the same Gini coefficient.  However, in cases where the two had the same color, they would not have been easily distinguished by the conventional means of looking at colors only.  The strength of the Gini coefficient, therefore, arises from its ability to break that degeneracy.  The significance of breaking this degeneracy is explored in Fig.~\ref{purities}.

In Fig.~\ref{purities}, we compare the relative abilities of different pairs of color and Gini coefficient information to distinguish between the 33 source galaxies in our sample.  For each combination of Ginis and/or colors, a plot like Fig.~\ref{Gini814vColor}. was constructed.  We defined regions of the Gini-color, Gini-Gini, or color-color space based on the outlines of each of the 33 clumps.  We then calculated the purity of each clump using the following process.   For the $i^{th}$ clump, we counted the number of points corresponding to galaxy $i$ inside the clump and divided by the sum of that number and the number of points from any galaxy $j\neq i$ inside that clump to determine a purity, where 1 is a clump that is perfectly separated from all of the others and lower values indicate more contamination from images of other galaxies.  We also compared methods using only a single Gini coefficient or only a single color.  In these cases, the regions are 1-dimensional and defined by the two images with the least and greatest Gini or color values.  Purities were calculated similarly for these 1-dimensional clumps.  The distribution purities arising from each combination of Gini-color, Gini-Gini, color-color, single Gini, or single color are shown in Fig.~\ref{purities}.  Red histograms correspond to methods of defining clumps that involve only Gini coefficient information, while blue histograms come from methods that use only color information, and purple histograms include both types of information.  The light red and light blue histograms denote methods that use only one piece of data (i.e., a single Gini or a single color) while the darker histograms use two pieces of data (two Ginis, two colors, or in the case of the purple histograms, one of each).

We find that methods of separating images into image families that include both a Gini coefficient and a color consistently yield noticeably higher purities than all other methods.  This means that the inclusion of morphological information from the Gini coefficient adds considerable power to image family identification methods above that would not be available from colors alone.  While we find that using the F606W Gini coefficient paired with the F606W-F814W color gave the highest purities the most often, it is unclear why these particular filters yielded the best results in this study.  It may be entirely due to the sharper in PSF in the F606W filter (and that the PSF is sharper for F814W images than for F160W images).  But it may also be because of some sort of morphological quality that is more present in the F606W band at low redshift or because the F606W and F814W filters capture the 4000\AA\ break in these low redshift galaxies (typically z$\sim$0.2-0.4), causing the F606W-F814W color to hold significant morphological information that would not be available in the F814W-F160W filter pair at these redshifts, but could be at higher redshifts.  To fully investigate this aspect of our result would require having a sample of highly spatially detailed images of galaxies with SEDs similar to those seen in the existing strong-lensing sample (i.e., with redshifts in the 1-3 range, or greater) which currently does not exist.  However, it may be possible to simulate such a sample using a code like GAMER \citep{Gamer} and a sample of SEDs from known moderate to high redshift galaxies that are more directly representative of typical lensed sources.  Gini analysis of well-studied strong-lensing clusters with robustly identified multiply-imaged families would allow an in-situ test of the applicability of this method. We will present this analysis in a future paper. 

Regardless of the reason for better results with some combinations of filters, it is clear that including the Gini coefficient from \textit{any} filter in attempts to identify image families is a substantial improvement over using colors alone.  And while degeneracies still remain even in the best filter combinations tested, it is possible that these degeneracies could be further broken by the inclusion of more Gini or color information (i.e., by using higher dimensional ``clumps") or by inclusion of some other measurement aside from these that is also preserved by gravitational lensing. This is exciting in the context of lens modeling, where accurate image family identification is required in order to make the best possible models of complex clusters like those in the Frontier Fields (e.g., \citealp{Jau14,Johnson14,Rich14,Grillo15,Ish15}).

\section{Conclusions}
We find that the Gini coefficient is likely to be an effective tool for the identification of image families in strong lensing systems with many images of an unknown number of source galaxies.  We have shown that the Gini coefficient, combined with a color, is capable of distinguishing between image families with effectiveness substantially greater than using one or two colors only.  This provides the additional benefit of minimizing the total number of observations that one needs to make in order to identify different image families.  Using two colors requires making observations in at least three filters, but using a Gini coefficient and a color requires only two.  This reduces the amount of observing time required to obtain reliable image family identifications by about a third, which is not insignificant given the tremendous demand for telescope resources.

Moreover, these results show that it is possible to automate a process--namely image family identification by image morphology--that is at present mostly a by-eye process requiring a considerable amount of researcher effort.  As we move into an era in which thousands of strong lenses will have imaging from space telescopes, such automation will be of great benefit.

\acknowledgments
Argonne National Laboratory's work was supported under the U.S. Department of Energy contract DE-AC02-06CH11357.

This work was supported in part by the Kavli Institute for Cosmological Physics at the University of Chicago through grant NSF PHY-1125897 and an endowment from the Kavli Foundation and its founder Fred Kavli, and by the Strategic Collaborative Initiative administered by the University of Chicago's Office of the Vice President for Research and for National Laboratories.

This research is also supported in part at the University of Chicago by the National Science Foundation under Grant PHY 08-22648 for the Physics Frontier Center ``Joint Institute for Nuclear Astrophysics" (JINA).

\clearpage

\end{document}